\documentclass[11pt]{article}
\usepackage{amsmath}
\usepackage{hyperref}
\pdfoutput=1

\begin{document}

\title{Scale Model Store Release from a Rectangular Cavity into Mach 1.5 Flow}

\author{Nathan E. Murray, Bernard Jansen \& Matthew Joachim\vspace{6pt}\\
National Center for Physical Acoustics\\
University of Mississippi\\
University, MS  38677\\
\\
Roger Birkbeck\vspace{6pt}\\
Combustion Research and Flow Technology Inc.\\
Pipersville, PA  18947}

\maketitle

\begin{abstract}
This fluid dynamics video shows the release of a $1/15^{\text{th}}$ scale model store into a Mach 1.5 free-stream flow from a rectangular cavity.  The images were acquired with a high-speed schlieren photography system using a Photron Ultima APX camera.  Acoustic measurements showed that the unsteady shear layer over the cavity had a prominent oscillation at approximately 810 Hz which corresponds to fluctuations seen in the cavity's leading edge shock structure in the video.  The video demonstrates the similarity in the time scales of the flow field dynamics and the residence time of the store in the shear layer.
\end{abstract}

\section{Experimental Setup \& Scaling}
Scale model free-drop store release experiments were conducted in the $12\times 12$ inch, Mach 5 capable, Tri-Sonic Wind Tunnel (TSWT) facility of the National Center for Physical Acoustics at the University of Mississippi.  A $1/15^{\text{th}}$ scale model store was ejected from a rectagular cavity into a Mach 1.5 free-stream flow using a novel pneumatic ejection system.  The rectangular cavity was 2 inches deep, 12 inches long, and 4 inches wide.  The model store was design using the ``light scaling'' method which allows for more realistic representation of pitch rates but under-represents the effects of gravity on the store trajectory.  As is common practice, this discrepancy was addressed by increasing the ejection force to around 38 Gs so that the scaled ejection velocity would correspond to full-scale.  This gives realistic time scales for the store trajectory in the initial phases of the release particularly as the store travels through the cavity shear layer.

\section{Video Details}
The store release event was captured using a high-speed schlieren photography system which utilized a continuous light source and a Photron Ultima APX camera.  The camera was set to capture images at $4000$ frames/sec with a shutter speed of $1/120000$ sec. (\href{http://ecommons.library.cornell.edu/bitstream/1813/13742/3/StoreReleaseSchlierenVideoLarge.mpg}{MPEG2 Video File} or \href{http://ecommons.library.cornell.edu/bitstream/1813/13742/2/StoreReleaseSchlierenVideoSmall.mpg}{MPEG1 Video File})

\section{Observations}
Acoustic measurements of both an empty and loaded cavity showed that the unsteady shear layer had a prominent oscillation at approximately 810 Hz.  This period of oscillation can be observed in the video to correspond to the fluctuations in the shock-waves both from the leading edge of the cavity and around the model store.  The video demonstrates the unsteady nature of the store release event and highlights the fact that the time scales of the flow field oscillations are on the same order of magnitude as the residence time of the store in the shear layer.  This suggests that the unsteady nature of the flow is a significant factor in determining the trajectory for stores released from an internal payload bay at supersonic speeds.

\end{document}